\documentclass[reprint,amsmath,amssymb,aps,superscriptaddress,onecolumn]{revtex4-1}

\usepackage{amssymb}
\usepackage{amsmath}
\usepackage{graphicx,color}
\usepackage{dcolumn}
\usepackage{bm}
\usepackage{epsfig,color}
\usepackage{epstopdf}

\begin{document}

\date{\today}

\title{Fermi-Pasta-Ulam chains with harmonic and anharmonic long-range
interactions}

\author{Gervais Nazaire Beukam Chendjou}
\affiliation{Fundamental Physics Laboratory: Group of Nonlinear
Physics and Complex Systems, Department of Physics, Faculty of
Science, University of Douala, Box 24157, Douala, Cameroon}
\affiliation{The Abdus Salam ICTP, Strada Costiera 11, I-34151
Trieste, Italy} \affiliation{SISSA, Via Bonomea 265, I-34136
Trieste, Italy} \affiliation{Istituto dei Sistemi Complessi, Consiglio Nazionale
delle Ricerche, via dei Taurini 19-00185 Roma, Italy}

\author{Jean Pierre Nguenang}
\affiliation{Fundamental Physics Laboratory: Group of Nonlinear
Physics and Complex Systems, Department of Physics, Faculty of
Science, University of Douala, Box 24157, Douala, Cameroon}
\affiliation{The Abdus Salam ICTP, Strada Costiera 11, I-34151
Trieste, Italy}

\author{Andrea Trombettoni}
\affiliation{CNR-IOM DEMOCRITOS Simulation Center, Via Bonomea 265,
I-34136 Trieste, Italy}
\affiliation{SISSA, Via Bonomea 265, I-34136
Trieste, Italy} \affiliation{INFN, Sezione di Trieste, I-34151
Trieste, Italy}

\author{Thierry Dauxois}
\affiliation{Univ. Lyon, ENS de Lyon, Univ. Claude Bernard, CNRS,
Laboratoire de Physique, F-69342 Lyon, France}

\author{Ramaz Khomeriki}
\affiliation{Department of Physics, Faculty of Exact and Natural
Sciences, Tbilisi State University, 0128 Tbilisi, Georgia}

\author{Stefano Ruffo}
\affiliation{SISSA, Via Bonomea 265, I-34136 Trieste, Italy}
\affiliation{INFN, Sezione di Trieste, I-34151 Trieste, Italy}
\affiliation{Istituto dei Sistemi Complessi, Consiglio Nazionale
delle Ricerche, via Madonna del Piano 10, I-50019 Sesto Fiorentino,
Italy}

\begin{abstract}
\vskip 2truecm
We study the dynamics of Fermi-Pasta-Ulam (FPU) chains 
with both harmonic and anharmonic power-law long-range
interactions. We show that the dynamics is described in the
continuum limit by a Generalized Fractional Boussinesq 
differential Equation (GFBE), whose derivation is performed in full
detail. We also discuss a version of the model where couplings are
alternating in sign.
\end{abstract}

\maketitle
\section{Introduction}\label{sec:I}

It is nowadays well recognized that the study of the
Fermi-Pasta-Ulam (FPU) model, first performed six decades ago, leads
to deep insights into the behavior of discrete nonlinear systems and
into fundamental problems of statistical mechanics
\cite{Chaos2005,Gallavotti2008}. The FPU model describes the
dynamics of an oscillator chain with nearest neighbor nonlinear
couplings among masses. At low energies, it exhibits the phenomenon 
of mode recurrence \cite{Fermi1955}, where the energy, 
initially fed into high wavelength modes, recursively returns 
to those modes. This remarkable effect was historically
tackled by considering the continuum limit, pioneered in the
classical paper by Zabusky and Kruskal \cite{Zabusky1965}. In
that limit one finds a Boussinesq equation \cite{davydov} and, after
a change of variables, a Korteweg-de Vries (KdV) equation
\cite{Zabusky1965} which gives an accurate description of waves and
localized solutions of the FPU model. Performing in a controlled way
the continuum limit for generalized FPU models is of paramount
importance, since it would allow one to construct solitonic solutions
and to determine the low-energy properties of these models.

Our goal here is to derive the effective equations describing the
dynamics of FPU models characterized by both harmonic and anharmonic
Long-Range Interactions (LRI). LRI have been intensively studied in
the last decades for a variety of physical systems \cite{campa14}
and considerable interest has been devoted to the study of power-law
LRI. Pioneering work by Dyson \cite{Dyson1969} has revealed that
the one-dimensional Ising ferromagnet with power-law couplings among
the spins displays a non trivial phase transition for values of the
power exponent $s$ in the range $1<s \le 2$.

Coupled oscillators with power-law LRI were also studied
\cite{Yu1997,Cuevas2002,Flach1998,Tarasov2006,Laskin2006,Korabel2007,Christodoulidi2014,Miloshevich2015,Miloshevich2017}.
The attention was mainly focused on long-range interactions in
DNA~\cite{Yu1997,Cuevas2002}, the existence of standing localized
solutions like breathers \cite{Flach1998}, deriving the continuum
counterpart of the discrete long-range models, which implies the use
of fractional derivatives \cite{Tarasov2006,Laskin2006,Korabel2007},
weakly chaotic \cite{Christodoulidi2014} and thermalization
\cite{Miloshevich2015} properties caused by the long-range character
of the interactions and the existence of solitons in a long-range
extension of the quartic FPU chain \cite{Miloshevich2017}.

In this paper, we will show that fractional differential equations describe
the continuum limit of the FPU model with both
harmonic and anharmonic power-law LRI.
Fractional calculus has gained considerable interest and importance
as an extension of differential equations with integer order
derivatives. These techniques are used for the investigation of
various problems in physics, engineering, life sciences and economy
\cite{Podluny1999,Kilbas2006,Malinowska2012,Gorenflo2001,Hilfer2000}.
The use of fractional derivatives may lead to an elegant and more
compact way of treating dynamical systems with non-local
interactions and/or couplings. Typical examples are 
fractional diffusion equations derived from the dynamics 
of oscillator chains 
using a hydrodynamic approach \cite{Beijeren,Spohn}. Also physical
phenomena with long memory \cite{Zaslavsky2002,Barkai2012} and
random displacements with space jumps of arbitrary length \cite{Metzler2000} 
have been considered.

Fractional derivatives are also used in condensed matter physics. We
can mention the recent studies of classical spin systems
\cite{defenu15,defenu16} and of fermionic quantum chains
\cite{lepori16,lepori17} with long-range couplings, which can be
described by effective field theories with a dispersion relation
associated in real space to a kinetic term with fractional
derivatives. Spatial fractional diffusion was also
observed in experimental cold atom physics \cite{Davidson} through
a mechanism induced by the interaction of atoms with the laser fields.

We here derive the Generalized Fractional Boussinesq 
differential Equation (GFBE), which describes the 
dynamics of the FPU model with LRI in both the harmonic and
anharmonic terms in the continuum
limit. 

The paper is organized as follows. In Section II we discuss the
derivation of the GFBE for the $\alpha$--FPU model. The
situation in which the couplings alternate in sign is presented
in Section III. In Section IV, we repeat all the derivations for the
$\beta$--FPU model. We discuss our results and perspectives in
Section V. In Appendix A we briefly review the basic definitions of
fractional calculus. In Appendixes B-D some derivations of lengthy
formulas needed for the continuum approximation of long-range FPU models
are presented. Appendix E is devoted to the case in
which cubic and quartic terms are both present.

\section{The $\alpha$--FPU model with power-law long-range interactions}
\label{alpha}

We consider in this Section a Hamiltonian where both the couplings
in the harmonic and the anharmonic terms of the $\alpha$--FPU model
have a power-law behavior with different exponents $s_1$ and
$s_2$, respectively
\begin{widetext}
\begin{equation}
H = \frac{1}{2M}\sum\limits_{n =  -
\infty }^{+\infty } {p_n^2  }   + \frac{\chi
}{2}\sum\limits_{\begin{array} {*{20}c}
   {n,m =  - \infty }  \\
   {m{\rm{ < }}n}  \\
\end{array}}^{+\infty } {\frac{{\left[ {u_n  - u_m } \right]^2 }}{{\left| {a\left( {n - m} \right)} \right|^{s_1} }}}
     + \frac{\gamma}{3}\sum\limits_{\begin{array}{*{20}c}{n,m =  - \infty }  \\
   {m{\rm{ < }}n}  \\
\end{array}}^{+\infty } {\frac{{\left[ {u_n  - u_m } \right]^3 }}{{\left| {a\left( {n - m} \right)} \right|^{s_2} }}},
 \label{alpha1}
\end{equation}
\end{widetext}
where $\chi$ and $\gamma$ are positive constants giving the strength
of the quadratic and cubic potentials, 
$n$ and $m$ stand for the
indices of the lattice sites, $a$ is the lattice spacing 
and $M$ is the mass (in the following $M \equiv 1$).

We choose the power-law decay exponents in the range $1<s_1, s_2<3$.
For smaller values the Hamiltonian diverges, while, when the powers tend
to infinity, we get back the conventional short-range $\alpha$--FPU model.
When both powers are finite and above $3$, the system becomes effectively
short-range. When $s_1$ (respectively $s_2$) is in the range $(1,3)$ and
$s_2$ (resp. $s_1$) is above $3$, then the effective continuum equation
displays a long-range behaviour in the anharmonic (harmonic) terms. 
The power $s_1=2$ (with $s_2 \to \infty$)
has been considered for crack front propagation along
disordered planes between solid blocks \cite{Laurson2013,Bonamy2008}
and for contact lines of liquid spreading on solid surfaces
\cite{Joanny1984}.

From Eq.~(\ref{alpha1}), we get the following equations of motion
\begin{widetext}
\begin{equation}
\ddot u_n  + \chi \sum\limits_{\begin{array}{*{20}c}
   {m =  - \infty }  \\
   {m \ne n}  \\
\end{array}}^{+\infty } {\frac{{u_n  - u_m }}{{\left| {a\left( {n - m} \right)} \right|^{s_1} }}}
              +  \gamma   \sum\limits_{\begin{array}{*{20}c}{m =  - \infty }  \\
   {m \ne n}  \\
\end{array}}^{+\infty } {\frac{{\left[ {u_n  - u_m } \right]^2 }}{{\left| {a\left( {n - m} \right)} \right|^{s_2} }}}{f_n,_m} = 0
  \label{alpha2}
\end{equation}
\end{widetext}
where
$$ {f_n,_m}  \equiv \left\{ {\begin{array}{*{20}c}
   +1, & {m < n}  \\
   -1, & {m > n}  \\
\end{array}} \right..$$

\begin{figure}[h!]
\includegraphics[width=3.3in,angle=0]{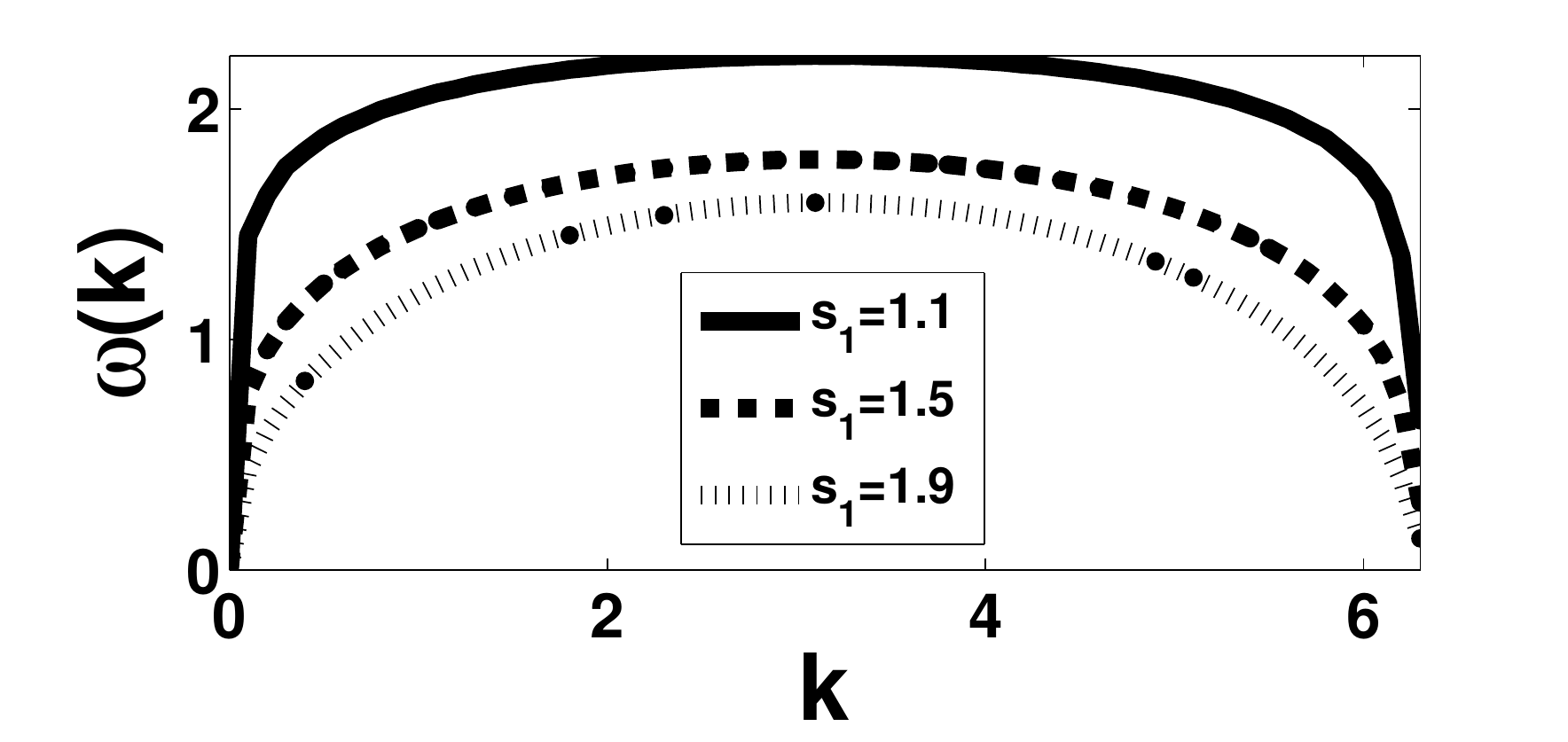}
\caption{Dispersion relation~(\ref{alpha4}) for 
$s_1=1.1, 1.5, 1.9$.}
\label{dispersionrelation}
\end{figure}

To understand the qualitative property of the wave equation, we
study the dispersion relation. We consider Eq.~(\ref{alpha2})  in
which we ignore the anharmonic terms.
Thus, Eq.~(\ref{alpha2}) admits the plane wave solution 
\begin{equation}
u_n \left( t \right) = u_0\, e^{i\left( {\omega t - kn } \right)} ,
\label{eq3}
\end{equation}
where $k$ is the wavenumber (we also set $x_n  \equiv na$).
The corresponding linear dispersion relation is
\begin{equation}
\omega^2(k) = 2\chi \sum\limits_{\ell= 1}^{\infty} {\frac{ {1 -\cos
\left( {k\ell} \right)} }{{\ell^{s_1} }}} ,
\label{alpha4}
\end{equation}
where from now on we put for simplicity $a \equiv 1$ 
(except when it is necessary for the analytical treatment).
Eq.~(\ref{alpha4}) is plotted in Fig.~(\ref{dispersionrelation}) for
three specific values of
$s_1$. It is straightforward to observe that the dispersion relation
diverges for $s_1<1$.

For small wavenumbers, $k\rightarrow 0$, $\omega(k)\propto |k|$ for
$s_1>3$ and the phase and group velocities are constant. If instead
$1<s_1<3$, $\omega(k)\propto |k|^{(s_1-1)/2}$ and the phase and
group velocities are given by $v_{ph} \propto | k|^{(s_1 - 3)/2}
{\rm{ }}$ and $v_g \propto | k |^{(s_1 - 3)/2} {\rm{ }}({s_1}-1)/2 =
v_{ph}({s_1}-1)/2$. They both diverge in the limit $k\rightarrow 0$.

Now we are ready for deriving the fractional partial differential
equation which describes Hamiltonian (\ref{alpha1}) in the long
wavelength, $k \to 0$, limit. Basic properties and definition of fractional
calculus are given for convenience in Appendix A. 
Let us start by defining the Fourier transform of $u_n\left( t \right)$ as

\begin{equation}
u_n \left( t \right) \equiv \frac{1}{{2\pi }}\int\limits_{ - \pi}^{
+\pi } {dk \, e^{ikn} \hat u\left( {k,t} \right)}
 \label{continuumalpha1}
\end{equation}
where 
\begin{equation}
\hat u\left( {k,t} \right) = \sum\limits_{n =  - \infty }^{ + \infty
} {e^{-ikn} } u_n \left( t \right)
 \label{continuumalpha2}.
\end{equation}

Substituing Eq.~(\ref{continuumalpha1}) into Eq.~(\ref{alpha2}), 
one gets separately for the sums in the latter the following expressions 
\begin{equation}
\sum\limits_{\scriptstyle m =  - \infty  \hfill \atop
  \scriptstyle m \ne n \hfill}^{+\infty } {\frac{{u_n  - u_m }}{{\left| {a\left( {n - m} \right)} \right|^{s_1} }}}  \approx  
  \frac{\pi }{{\Gamma \left( s \right)\sin \frac{{{s_1} - 1}}{2}\pi }}\frac{1}{{2\pi }}\int\limits_{ - \infty }^{ +\infty } {dp\,\,e^{ipx} \left| p \right|^{s_1 - 1} \hat u\left( {p,t} \right)}
 \label{continuumalpha1_14} ,
\end{equation}

\begin{equation}
\sum\limits_{\begin{array}{*{20}c}
   {m =  - \infty }  \\
   {m \ne n}  \\
 \end{array} }^{ + \infty } {\frac{{\left( {u_n - u_m } \right)^2 }}
{{\left| {a\left( {n - m} \right)} \right|^{s_2} }}} f_{n,m} \approx
\frac{{u\left( {x,t} \right)}} {{2\pi }}\int\limits_{ - \infty }^{ +
\infty } {dp\,\,e^{ipx}
\frac{{\pi}} {{\Gamma \left( s_2 \right)\sin \left( {s_2\pi } \right)}} 
\left(
{\left( { - ip} \right)^{s_2 - 1}  - \left( {ip} \right)^{s_2 - 1} }
\right)} \hat u\left( {p,t} \right) ,
 \label{continuumalpha2_10}
\end{equation}
where $p \equiv k/a$ and $u_n(t) \equiv a\,u(x,t)$ as $a \to 0$. 
As detailed in Appendix B, in deriving 
Eq.~(\ref{continuumalpha2_10}) for the nonlinear term of the FPU,  
in the lattice, before taking the continuum limit, one gets sums of the 
type $\sum_{n'} u_{n\pm n'} \cdots$. 
Doing the approximation that in such sums the $u_{n \pm n'}$'s are slowly varying 
in space in the continuum limit, one brings the terms $u_{n\pm n'}$ outside 
the sums over $n'$, and Eq.~(\ref{continuumalpha2_10}) is found. 

In the continuum limit, the Fourier sum becomes the Fourier transform, 
hence

\begin{equation}
\hat u\left( {p,t} \right) = \int\limits_{ - \infty
}^{ + \infty } {dx \,e^{ - ipx}\, u\left( {x,t} \right)} ,\,\,\,\,\,\,\,\,\,\,\,\,\,\,\,\,\,\,\,\,\,\,\,\,\,
 u\left( {x,t} \right) = \frac{1}{{2\pi }}\int\limits_{ - \infty
}^{ + \infty } {dp \,e^{ ipx}\,\hat u\left( {p,t} \right)} .
\label{continuumalpha5}
\end{equation}

We now introduce the Fourier transformation of fractional differentiation of order $\alpha$
via the relations
\begin{equation}
D_{x^ +  }^\alpha  u\left( {x,t} \right)  \equiv
\frac{{\partial ^\alpha  }}{{\partial x^\alpha  }}u\left( {x,t}
\right)= \frac{1}{{2\pi
}}\int\limits_{ - \infty }^{ + \infty } {dp\left( { + ip}
\right)^\alpha e^{ipx}\, \hat u\left( {p,t} \right)}\,\,
 \label{continuumalpha7}
\end{equation}
and
\begin{equation}
D_{x^ -  }^\alpha  u\left( {x,t} \right) = \frac{1}{{2\pi
}}\int\limits_{ - \infty }^{ + \infty } {dp\left( { - ip}
\right)^\alpha e^{ipx}\, \hat u\left( {p,t} \right)}.
 \label{continuumalpha8}
 \end{equation}
The Fourier transform involving the absolute value of momentum
$|p|^\alpha$ is expressed by a Riesz derivative in real space as
\begin{equation}
-\frac{{\partial ^\alpha  }}{{\partial \left| x \right|^\alpha
}}u\left( {x,t} \right) = \frac{1}{{2\pi }}\int\limits_{ - \infty
}^{ + \infty } {dp\left| p \right|^\alpha  \hat u\left( {p,t}
\right)} e^{ipx}, \label{continuumalpha9}
\end{equation}
and, using Eqs.~(\ref{continuumalpha7}) and~(\ref{continuumalpha8}), one gets
\begin{equation}
\frac{{\partial ^\alpha  }}{{\partial \left| x \right|^\alpha}}u
\left( {x,t} \right) = -\frac{1 }{2{\cos
\frac{{{\alpha\pi}}}{2}}}[D_{x^ + }^\alpha + D_{x^ -
}^\alpha]u\left( {x,t} \right)
 \label{continuumalpha10}.
\end{equation}

Combining Eqs.~(\ref{continuumalpha1_14}) and (\ref{continuumalpha2_10})
with Eqs.~(\ref{continuumalpha7}), (\ref{continuumalpha8}) and (\ref{continuumalpha10}), one ends up with

\begin{equation}
 \sum\limits_{\begin{array}{*{20}c}
   {m =  - \infty {\text{  }}}  \\
   {m \ne n{\text{ }}}  \\
 \end{array} }^{ + \infty } {\frac{{u_n  - u_m }}
{{\left| {a\left( {n - m} \right)} \right|^{s_1} }}}  \approx - \frac{\pi }
{{\Gamma \left( s_1 \right)\sin \frac{{s_1 - 1}} {2}\pi
}}\frac{{\partial ^{s_1 - 1} }} {{\partial \left| x \right|^{s_1 - 1}
}}u\left( {x,t} \right)
\label{continuumalpha1_15},
\end{equation}

\begin{widetext}
\begin{equation}
 \sum\limits_{\begin{array}{*{20}c}
   {m =  - \infty }  \\
   {m \ne n}  \\
 \end{array} }^{ + \infty } {\frac{{\left( {u_n  - u_m } \right)^2 }}
{{\left| {a\left( {n - m} \right)} \right|^{s_2} }}} f_{n,m}  \approx
\frac{{\pi}} {{\Gamma \left( s_2 \right)\sin \left( {s_2\pi } \right)}} u\left( {x,t} \right)\left[
{D_{x^ -  }^{s_2 - 1}  - D_{x^ +  }^{s_2 - 1} } \right]u\left( {x,t}\right) .
 \label{continuumalpha2_11}
\end{equation}
\end{widetext}

Finally, Eqs.~(\ref{continuumalpha2_11}), 
(\ref{continuumalpha1_15}), (\ref{continuumalpha8}), 
(\ref{continuumalpha7}) and (\ref{alpha2}) combine
into the following Generalized Fractional Boussinesq Equation
(GFBE)
\begin{widetext}
\begin{equation}
\frac{{\partial ^2 u\left( {x,t} \right)}} {{\partial t^2 }}\,\, -\,\,
g_{s_1  - 1} \frac{{\partial ^{s_1  - 1} u\left( {x,t} \right)}}
{{\partial \left| x \right|^{s_1  - 1} }} -  h_{s_2  - 1} u\left(
{x,t} \right)\left[ {D_{x^ -  }^{s_2  - 1}  - D_{x^ +  }^{s_2  - 1}
} \right]u\left( {x,t} \right) = 0 ,
 \label{continuumlimitalpha16}
\end{equation}
\end{widetext}
where the constants $g_{{s_1} - 1} $ and $ h_{{s_2} - 1} $ are given by
\begin{equation}
g_{{s_1} - 1}  = \frac{{\chi \pi }}{{\Gamma \left({s_1}\right)\sin
\left( {\frac{{{s_1} - 1}}{2}\pi} \right)}}, \,\,\,\,\,\, h_{{s_2}-1}  = - \frac{{{\gamma  }\pi
}}{{\Gamma\left( {s_2} \right)\sin {\left({s_2}\pi\right)}}}.
 \label{continuumlimitalpha17} 
\end{equation}
Eq.~(\ref{continuumlimitalpha16}) is the main result of the paper. We observe that the GFBE is not defined for integer order derivatives.

\section{$\alpha$--FPU model with alternating masses and
interactions} \label{alternatingalpha}

The main goal of the paper is to establish that general FPU chains
with LRI are mapped onto fractional equations of motion and we aim at
illustrating it in a variety of models. To investigate how general is
this mapping, we consider in this Section an $\alpha$-FPU model with
alternating  signs in kinetic and interacting terms. Despite the
physical realization of both alternating masses and interaction
terms is certainly not easily implementable, alternating/varying
interactions are rather common and one can think to implement
alternating in sign hopping coefficients in ultracold chains using a
variation of well-known shaking techniques \cite{eckardt05}. The
corresponding Hamiltonian is written as


\begin{widetext}
\begin{equation}
H = \frac{1}{2}\sum\limits_{n =  -
\infty }^{+\infty } {p_n^2 } \left( { - 1} \right)^n  +
\frac{\chi }{4}\sum\limits_{\begin{array}{*{20}c}
   {n,m =  - \infty }  \\
   {m{\rm{ \ne }}n}  \\
\end{array}}^{+\infty } {\frac{{\left[ {u_n - u_m } \right]^2 \left( {- 1} \right)^n }}{{\left| {a\left( {n - m} \right)} \right|^s }}}
    + \frac{\gamma }{6}\sum\limits_{\begin{array}{*{20}c}
   {n,m =  - \infty }  \\
   {m{\rm{ \ne }}n}  \\
\end{array}}^{+\infty } {\frac{{\left[ {u_n  - u_m } \right]^3 \left( { - 1} \right)^n }}{{\left| {a\left( {n - m} \right)} \right|^s }} f_{n,m}}
 \label{alternatingalpha1}
 \end{equation}
\end{widetext}
where $\chi$ and $\gamma$ are positive constants. Here again the
exponent is chosen in the range of $1 < s <3$. Applying the
Hamiltonian formalism to Eq.~(\ref{alternatingalpha1}), one gets the following
equation on the lattice
\begin{equation}
\ddot u_n  +  \chi  \sum\limits_{\begin{array}{*{20}c}
   {m =  - \infty }  \\
   {m \ne n}  \\
\end{array}}^{+\infty } {\frac{{u_n  - u_m }}{{\left| {a\left( {n - m} \right)} \right|^s }}}
              +  \gamma  \sum\limits_{\begin{array}{*{20}c}{m =  - \infty }  \\
   {m \ne n}  \\
\end{array}}^{+\infty } {\frac{{\left[ {u_n  - u_m } \right]^2 }}{{\left| {a\left( {n - m} \right)} \right|^s }}}  = 0.
  \label{alternatingalpha2}
\end{equation}
We can obtain a plane wave solution as
\begin{equation}
u_n \left( t \right) = u_0\,  e^{i\left( {\omega t - k n } \right)},
\label{alternatingalpha3}
\end{equation}
with the dispersion relation of the fractional wave
equation given in Eq.~(\ref{alpha4}).

To derive the continuum equation describing the system while using
the long wavelength limit for the corresponding lattice-field model,
we use the expressions of the Fourier series 
defined in Eqs.~(\ref{continuumalpha1}) and~(\ref{continuumalpha2}). 
The third term of Eq.~(\ref{alternatingalpha2}) is therefore  transformed (see Appendix C)
 into
 
\begin{widetext}
\begin{equation}
\sum\limits_{\begin{array}{*{20}c}
   {m =  - \infty }  \\
   {m \ne n}  \\
 \end{array} }^{m =  + \infty } {\frac{{\left( {u_n  - u_m } \right)^2 }}
{{\left| {a\left( {n - m} \right)} \right|^s }}}  \approx {
\frac{{\pi }} {{\Gamma \left( s
\right)\sin \left( {\frac{{s - 1}} {2}\pi } \right)}}}\frac{{u\left(
{x,t} \right)}} {{2\pi }}\int\limits_{ - \infty }^{ + \infty } {dp}
\left| p \right|^{s - 1} e^{ipx} \hat u\left( {p,t} \right).
 \label{alteratingalpha12}
\end{equation}
\end{widetext}

Using Riesz fractional derivative, definitions Eqs.~(\ref{continuumalpha9}) and (\ref{continuumalpha10}), we obtain

\begin{widetext}
\begin{equation}
 \sum\limits_{\begin{array}{*{20}c}
   {m =  - \infty }  \\
   {m \ne n}  \\
 \end{array} }^{+ \infty } {\frac{{\left( {u_n  - u_m } \right)^2 }}
{{\left| {a\left( {n - m} \right)} \right|^s }}}  \approx  
-\frac{{\pi }} {{\Gamma \left( s
\right)\sin \left( {\frac{{s - 1}} {2}\pi } \right)}} u\left( {x,t}
\right)\frac{{\partial ^{s - 1} }} {{\partial \left| x \right|^{s -
1} }}u\left( {x,t} \right).
 \label{alteratingalpha13}
\end{equation}
\end{widetext}
Finally, Eqs.~(\ref{continuumalpha1_15}), (\ref{alteratingalpha13}) and (\ref{alternatingalpha2}) can be combined
into the following GFBE 
\begin{widetext}
\begin{equation}
\frac{{\partial ^2 }}{{\partial t^2 }}u\left( {x,t} \right) \,\, - \,\,g_{{s}
- 1} \frac{{\partial ^{{s} - 1} }}{{\partial \left| x \right|^{{s}-
1} }}u\left( {x,t} \right) \,\, - \,\, k_{{s} - 1} u\left( {x,t}
\right)\frac{{\partial ^{{s} - 1} }}{{\partial \left| x \right|^{{s}
- 1} }}u\left( {x,t} \right) = 0,
 \label{alternatingalpha11}
\end{equation}
\end{widetext}
with
\begin{widetext}
\begin{equation}
k_{{s} - 1}  = \frac{{\gamma}  \pi}{{\Gamma \left( {s} \right)\sin \left(\frac{{{s-1}}}{2}\pi \right) }}.
 \label{alternatingalpha12}
\end{equation}
\end{widetext}
Despite the presence of the alternating terms $(-1)^n$ in
Eq.~(\ref{alternatingalpha1}), one finds again a fractional
differential equation, even simpler than Eq.~(\ref{alpha1}). One can
also write Eq.~(\ref{alternatingalpha11}) as
\begin{equation}
u_{tt}  \,\, -  \,\, g _{s - 1} D_{\left| x \right|}^\alpha  u   \,\,- \,\, k _{s - 1} u D_{\left| x \right|}^\alpha  u = 0
 \label{alternatingalpha111}
\end{equation}
where $\alpha  = s - 1$.

\section{The $\beta$--FPU model with power-law long-range interactions}
\label{Beta}

In this Section we study the effect of LRI both in the harmonic and
anharmonic terms of an extended $\beta$--FPU Hamiltonian model. The
main difference with the previous Sections is that now the
interaction is quartic instead of cubic - for the rest our goal is
to parallel the results presented in the previous Sections. The model
reads

\begin{widetext}
\begin{equation}
H  = \frac{1}{2}\sum\limits_{n =  -
\infty }^{+\infty } {p_n^2 }  + \frac{\chi
}{4}\sum\limits_{\begin{array}{*{20}c}
   {n,m =  - \infty }  \\
   {m \ne n}  \\
\end{array}}^{+\infty } {\frac{{\left[ {u_n  - u_m } \right]^2 }}{{\left| {a\left( {n - m} \right)} \right|^{s_1} }} + \frac{\gamma }{8}\sum\limits_{\begin{array}{*{20}c}
   {n,m =  - \infty }  \\
   {m \ne n}  \\
\end{array}}^{+\infty } {\frac{{\left[ {u_n  - u_m } \right]^4 }}{{\left| {a\left( {n - m} \right)} \right|^{s_2} }}} }
 \label{beta1}
\end{equation}
\end{widetext}
where $\chi$ and $\gamma$ are positive constants.  
Hamiltonian
(\ref{beta1}) is referred to as the extended $\beta$--FPU model.
Again we choose the parameters describing the order of the
fractional space derivative in the range $1 < {s_1} ,
{s_2} < 3$. The equations of motion read
\begin{widetext}
\begin{equation}
\ddot u_n  + \chi \sum\limits_{\begin{array}{*{20}c}
   {m =  - \infty {\rm{ }}}  \\
   {m \ne n}  \\
\end{array}}^{+\infty } {\frac{{u_n  - u_m }}{{\left| {a\left( {n - m} \right)} \right|^{s_1} }}}  + \gamma \sum\limits_{\begin{array}{*{20}c}
   {m =  - \infty }  \\
   {m \ne n{\rm{ }}}  \\
\end{array}}^{+\infty } {\frac{{\left[ {u_n  - u_m } \right]^3 }}{{\left| {a\left( {n - m} \right)} \right|^{s_2} }}}  = 0
 \label{beta2}
\end{equation}
\end{widetext}

Doing the same analytical calculation illustrated in detail in the
previous Sections and Appendices B and C, we can derive the continuum
equation describing the macroscopic system within the
long-wavelength limit framework. The third term of Eq.~(\ref{beta2})
can be rewritten as (see for more details Appendix D)

\begin{widetext}
\begin{equation}
\sum\limits_{\begin{array}{*{20}c}
   {m =  - \infty }  \\
   {m \ne n}  \\
 \end{array} }^{m =  + \infty } {\frac{{\left( {u_n  - u_m } \right)^3 }}
{{\left| {a\left( {n - m} \right)} \right|^{s_2} }}}  \approx {
\frac{{\pi }} {{\Gamma \left( s_2 \right)\sin \left( {\frac{{{s_2} - 1}} {2}\pi }
\right)}}}\frac{{u^2 \left( {x,t} \right)}} {{2\pi }}\int\limits_{ -
\infty }^{ + \infty } {dp} \left| p \right|^{s_2 - 1} e^{ipx} \hat
u\left( {p,t} \right).
 \label{beta10}
\end{equation}
\end{widetext}

Using Riesz fractional derivative definitions,
Eqs.~(\ref{continuumalpha9}) and (\ref{continuumalpha10}), we
obtain

\begin{widetext}
\begin{equation}
 \sum\limits_{\begin{array}{*{20}c}
   {m =  - \infty }  \\
   {m \ne n}  \\
 \end{array} }^{m =  + \infty } {\frac{{\left( {u_n  - u_m } \right)^3 }}
{{\left| {a\left( {n - m} \right)} \right|^{s_2} }}}  \approx 
-\frac{{\pi }} {{\Gamma \left( s_2 \right)\sin \left( {\frac{{{s_2} - 1}} {2}\pi }\right)}} u^2
\left( {x,t} \right)\frac{{\partial ^{s_2 - 1} }} {{\partial \left| x
\right|^{s_2 - 1} }}u\left( {x,t} \right)
 \label{beta11}.
\end{equation}
\end{widetext}

Substituting Eqs.~(\ref{continuumalpha1_15}) and (\ref{beta11}) into
Eq.~(\ref{beta2})  it follows that
Eq.~(\ref{beta2}) is reduced to the following GFBE
\begin{widetext}
\begin{equation}
\frac{{\partial ^2 }}{{\partial t^2 }}u\left( {x,t} \right) \,\, - \,\,
g_{{s_1} - 1} \frac{{\partial ^{{s_1} - 1} }}{{\partial \left| x
\right|^{{s_1} - 1} }}u\left( {x,t} \right) \,\, - \,\, j_{{s_2} - 1} u^2
\left( {x,t} \right)\frac{{\partial ^{{s_2} - 1} }}{{\partial \left|
x \right|^{{s_2} - 1} }}u\left( {x,t} \right) = 0,
 \label{beta12}
\end{equation}
\end{widetext}

with

\begin{widetext}
\begin{equation}
 j_{{s_2} - 1}  = \frac{{\pi\gamma}} {{\Gamma \left( s_2 \right)\sin \left( {\frac{{{s_2} - 1}} {2}\pi }
\right)}}.
\label{beta13}
\end{equation}
\end{widetext}
Finally, we report in Appendix E the results for
the $\alpha$+$\beta$--FPU model in which cubic and quartic terms
appear simultaneously.

\section{Conclusions and perspectives}
\label{conclusions}

Starting from the lattice dynamics of the FPU model with both
harmonic and anharmonic long-range power law couplings, we have
derived in the continuum limit a Generalized Fractional Boussinesq
Equation (GFBE)  as it is usually done for short-range
model \cite{davydov,Zabusky1965}. We have performed the analytical
derivations by two different methods: i) using Riesz derivative and
Hurwitz formula of fractional calculus, ii) performing a direct
analysis of the Fourier spectrum in the $k\to 0$ limit. We also dealt
with a variant of the model where masses and couplings are alternating
in sign.

In general, the presence of long-range couplings in the FPU model is
reflected into the appearance of nonlocal terms in the continuum
equations, the nonlocality being mathematically represented by
fractional derivatives. When the power of the couplings
tends to infinity, the interactions become short-range, nonlocality
is thus removed and fractional derivatives convert into ordinary partial
derivatives. In this paper, the fractional derivatives are mainly
considered in the  Riesz sense.

Our systematic formulation of mechanics based on fractional
derivatives can be used to develop models of biological systems in
which fractional power-law interaction are essential elements of
biological phenomena (for instance anomalous diffusion in cell
biology).  In the area of physics, fractional space derivatives are
used to model anomalous diffusion or dispersion, where a particle
spreads at a rate inconsistent with classical Brownian motion 
\cite{Metzler2000}. In particular, the Riesz fractional
derivative includes a left Riemann-Liouville derivative and a right
Riemann-Liouville derivative that allows one to model flow regimes where 
impacts occur from either side of the domain \cite{Zaslavsky2002}.

In the future work, we plan to study solutions of the GFBE, 
Eqs.~(\ref{continuumlimitalpha16}), (\ref{alternatingalpha11}),  
(\ref{beta12}) and (\ref{alpha+beta3}). Extended wave solutions could be derived using the homotopy analysis method \cite{Liao2003}  
or the rotating wave approximation \cite{Orszag}. Also localized 
solutions may exist, but for what we could preliminarily 
find numerically \cite{progress}, 
they seem to be unstable for the lattice equations. 
All these issues of correspondence between solutions of GFBE and 
FPU lattice equations will be the subject of forthcoming studies.

The effect of different boundary
conditions should also be analyzed in much detail due to its importance
for models with long-range interactions. Another challenge for the
future would be to find the correct transformation of space and time
variable which would allow one to derive from the GFBE the
generalized KdV equation valid for long-range interactions.

\vspace{1cm}

\section*{Acknowledgments}
Discussions with G. Gori, N. Defenu and Y. Zhou are acknowledged. 
We are also grateful to M. Gallone for discussions and for pointing 
out the errouneous application of the 
Fourier transform in the non linear 
terms of the equations of motion.
The support from the OFID Postgraduate Fellowship Programme at ICTP
and from the ICTP/IAEA Sandwich Training Educational Programme is
gratefully acknowledged. The work is supported in part by travel
grants from PICS CNRS (Grant No 04/01) and CNR (Grant No 04/24).
R. Kh. is funded by SRNSF (Grant Nos 216662,
FR/25/6-100/1404/01 and STCU-2016-03) and by STCU (Grant No 6303).

 \newpage

 \bigskip

\appendix

\section{Basic definitions of fractional calculus}
\label{AppendixA}

In this Appendix, we briefly review basic notions of
fractional differentiation and
integration. Fractional differentiation is an extension of the order
of differentiation from an integer to a real number. 
We also introduce Riesz derivatives,
which are the most useful representation applicable to a 
model with long-range interaction \cite{Ishiwata2012}. In addition,
we present fractional integration, which is the inverse operation of
fractional differentiation.\\

\textbf{Definition 1.}  A real function $u(x)$, $x>0$, is said to be
in the space $C_\mu$, $\mu\in\textbf{R}$, if there exists a real
number $p>\mu$, such that $u(x)=x^{p}u_1(x)$, where
$u_1(x)\in{C(0,\infty)}$, and it is said to be in the space
${C}_{\mu}^{n}$ if and only if 
$d^nu(x)/dx^n \in C_\mu, n \in\textbf{N}$ \cite{Podluny1999}.\\

\textbf{Definition 2.}  The definition of a fractional derivative of
order $\alpha$ $(\alpha>0)$ for $u(x) \in C_\mu$ ($\mu \ge -1$) 
with respect to $x$ is
formulated in the following two ways
\begin{equation}
D_{x^{+}}^{\alpha}u(x)=\frac{1}{\Gamma(n-\alpha)}\left({\frac{d}{dx}}\right)^{n}\int\limits_{
-\infty}^{ x} dy\,\,u(y)\,{(x-y)^{n-\alpha-1}},
 \label{AppendixA1}
\end{equation}

\begin{equation}
D_{x^{-}}^{\alpha}u(x)=\frac{1}{\Gamma(n-\alpha)}\left({-\frac{d}{dx}}\right)^{n}\int\limits_{
x}^{ +\infty} dy\,\,u(y)\,{(y-x)^{n-\alpha-1}}.
 \label{AppendixA2}
\end{equation}

In this formula, the integer $n$ is chosen for a given real number
$\alpha$ such that $n-1\leqslant\alpha < n$, and $\Gamma(n-\alpha)$
denotes Euler's gamma function. When an order $\alpha$ is an integer
$n$, a fractional derivative Eq.~(\ref{AppendixA1}) and
Eq.~(\ref{AppendixA2}) is reduced to standard derivatives of integer
order

\begin{equation}
D_{x^{+}}^{n}=\left({\frac{d}{dx}}\right)^{n},\,\,\,\,\,\,D_{x^{-}}^{n}=\left({-\frac{d}{dx}}\right)^{n}.
 \label{AppendixA3}
\end{equation}

Therefore, $D_{x^{+}}^{n}= (-1)^{n} D_{x^{-}}^{n}$, but in general
$D_{x^{+}}^{\alpha}\ne (-1)^{\alpha} D_{x^{-}}^{\alpha}$; \,\,see
Ref. \cite{Ishiwata2012} and

\begin{equation}
 D_{x^{+}}^{\alpha}x^{\gamma}=\frac{\Gamma(\gamma+1)}
{\Gamma(\gamma-\alpha+1)}x^{\gamma-\alpha},\,\,\,\,\,\,\,\,\, \mbox{for}\,\,\,\,\,\,\,\,\,\,\,\, x>0\,\,\,\mbox{and}\,\,\,\, \gamma \ge 0 .
 \label{AppendixA4}
\end{equation}

The formulation of fractional differentiation \cite{Podluny1999} includes 
 integration of order $\alpha$ for $u(x)$, as defined in the
following two ways

\begin{equation}
I_{x^{+}}^{\alpha}u(x)=\frac{1}{\Gamma(\alpha)}\int\limits_{
-\infty}^{ x} {dy\,u(y)\,(x-y)^{\alpha-1}},
 \label{AppendixA5}
\end{equation}

\begin{equation}
I_{x^{-}}^{\alpha}u(x)=\frac{1}{\Gamma(\alpha)}\int\limits_{
x}^{ +\infty} {dy\,u(y)\,(y-x)^{\alpha-1}}.
 \label{AppendixA6}
\end{equation}

\begin{equation}
I^{0}u(x)=u(x).
\label{AppendixA7}
\end{equation}
Fractional differentiation of order $\alpha$ is the inverse operation of fractional integration of order $\alpha$
\begin{equation}
D_{\mu^{+}}^{\alpha}I_{\mu^{+}}^{\alpha}u(x)=u(x),\,\,\,\,\,D_{\mu^{-}}^{\alpha}I_{\mu^{-}}^{\alpha}u(x)=u(x) \label{AppendixA8}.
\end{equation}
If we assume $u(x)$ to be a regular function in $-\infty<x<\infty$, an integer derivative and a fractional derivative are commutable as follows
\begin{equation}
D_{\mu^{+}}^{\alpha}\left(\partial_{\nu}u(x)\right)=\partial_{\nu}\left(D_{\mu^{+}}u(x)\right),\,\,\,\,\,D_{\mu^{-}}^{\alpha}\left(\partial_{\nu}u(x)\right)=\partial_{\nu}\left(D_{\mu^{-}}u(x)\right).
\label{AppendixA9}
\end{equation}

However, the commutation relation between two fractional derivatives is not satisfied in general
\begin{equation}
D_{\mu^{+}}^{\alpha}\left(D_{\mu^{-}}^{\beta}u\right)\ne D_{\mu^{-}}^{\beta}\left(D_{\mu^{+}}^{\alpha}u\right)
\label{AppendixA10}.
\end{equation}

\textbf{Definition 3.} The Caputo fractional derivative
$D^{\alpha}$ of $u(x)$ is defined~\cite{Podluny1999} as
\begin{equation}
D^{\alpha}u(x)=\frac{1}{\Gamma(n-\alpha)}\int\limits_{ 0}^{ x}
{(x-\xi)^{n-\alpha-1}u^{(n)}(\xi)d\xi}\,\,\,\,\,\,(\alpha>0),
  \label{AppendixA11}
\end{equation}
for $n-1<\alpha\leqslant{n}$, $n\in\textbf{N}$, $x>0$, $u\in{C}_{-1}^{n}$.
The following are two basic properties of the Caputo fractional derivative:\\
$(1)$ Let  $u\in{C}_{-1}^{n}$, $n\in\textbf{N}$. Then $D^{\alpha}u$, $0\leqslant\alpha\leqslant{n}$ is well defined and  $D^{\alpha}u\in{C}_{-1}$.\\
 $(2)$ Let $n-1<\alpha\leqslant{n}$, $n\in\textbf{N}$ and
$u\in{C}_{\lambda}^{n}$, $\lambda\geqslant-1$.

Then one has
\begin{equation}
(I^{\alpha}D^{\alpha})u(x)=u(x)-\sum\limits_{\scriptstyle k= 0}^{
n-1 }{u^{k}(0^{+})}\frac{x}{k!},\,\,\,\,\,\, (x>0).
 \label{AppendixA12}
\end{equation}

In this paper, only real and positive values of $\alpha$ have been considered.
Similar to integer-order differentiation, the Caputo fractional
differentiation is a linear operation
\begin{equation}
D^{\alpha}(\lambda{f(x)}+\mu{g(x)})=\lambda{D^{\alpha}{f(x)}}+\mu{D^{\alpha}{g(x)}},\label{eq141}
\end{equation}
where $\lambda$, $\mu$ are constants.\\

\textbf{Definition 4.}  A generalization of the classical Leibniz rule
\begin{equation}
D^{n}({f}{g})= \sum\limits_{\scriptstyle k=
0}^{\infty}\left(\begin{array}{c}n
\\k\end{array}\right) \left(f^{n-k}\right) g^k
\label{Appendix0A13}
\end{equation}
from integer $n$ to fractional $\alpha$ contains an infinite series
\begin{equation}
D^{\alpha}({f}{g})(x) = \sum\limits_{\scriptstyle k=
0}^{\infty}\left(\begin{array}{c}\alpha
\\k\end{array}\right) \left(D^{\alpha-k}_{x} {f} \right)(x) D^k_x \,g
\label{AppendixA13},
\end{equation}
with $f(\xi)$ continuous in $[0,\,x]$ and $g(\xi)$ having $(n+1)$
continuous derivatives in $[0,\,x]$. The sum is infinite and contains integrals of fractional order (for $k> [\alpha]+1$).\\

\textbf{Definition 5.}  Let $n$ be the smallest integer that exceeds $\alpha$, then the Caputo fractional derivative
operator of order $\alpha>0$ with respect to $x_\mu$, is formulated~\cite{Podluny1999}
as
 \begin{equation}
D^{\alpha}_{x_\mu+}u(x)=\frac{1}{\Gamma(n-\alpha)}\int\limits_{
-\infty}^{ x_\mu}dy\frac{\partial^n}{\partial{y}^n}
{u(x_1,...,y,x_{\mu_{+1}},...)}{(x_\mu-y)^{n-\alpha-1}}
\label{AppendixA14}
 \end{equation}
and
 \begin{equation}
D^{\alpha}_{x_\mu-}u(x)=\frac{(-1)^n}{\Gamma(n-\alpha)}\int\limits_{
x_\mu}^{ + \infty}dy\frac{\partial^n}{\partial{y}^n}
{u(x_1,...,y,x_{\mu_{+1}},...)}{(y-x_\mu)^{n-\alpha-1}}
\label{AppendixA15}
 \end{equation}
 if $n-1<\alpha<n$. Of course $D^{\alpha}_{x}u(x)=\frac{\partial^n{u(x)}}{\partial{x^n}}$ if
$\alpha\in\textbf{N}$. The corresponding Fourier transformation of
order $\alpha$ is defined as \cite{Ishiwata2012}
\begin{equation}
 D_{x^ +  }^\alpha  u\left( {x,t} \right) = \frac{1}{{2\pi
}}\int\limits_{ - \infty }^{ + \infty } {dp\left( { + ip}
\right)^\alpha  \hat u\left( {p,t} \right)} e^{ipx}
\label{AppendixA17}
\end{equation}
\begin{equation}
D_{x^ -  }^\alpha  u\left( {x,t} \right) = \frac{1}{{2\pi
}}\int\limits_{ - \infty }^{ + \infty } {dp\left( { - ip}
\right)^\alpha  \hat u\left( {p,t} \right)} e^{ipx},
 \label{AppendixA18}
\end{equation}
where
 \begin{equation}
\hat u\left( {p,t} \right) = \frac{1}{{2\pi }}\int\limits_{ - \infty
}^{ + \infty } {dp\,u\left( {x,t} \right)} e^{ - ipx}.
 \label{AppendixA19}
\end{equation}
Riesz derivatives can be defined as follows, except when $\alpha$ is
an odd number \cite{Yang2010}
\begin{equation}
\frac{{\partial ^\alpha  }}{{\partial \left| x \right|^\alpha
}}u\left( {x,t} \right) = -\frac{1 }{2{\cos
\frac{{{\alpha\pi}}}{2}}}[D_{x^ +  }^\alpha + D_{x^ -
}^\alpha]u\left( {x,t} \right).
 \label{AppendixA20}
\end{equation}
The above derivative is singular at $\alpha=1,3,5,...$. Riesz
derivatives have symmetry with respect to the transformation $x$ $\to$
$-x$. When $u(x)$ is a regular function in $-\infty<x<\infty$, the two formulations provided by Riemann-Liouville and Caputo are equivalent. We note that the commutation relation between the two fractional derivatives defined by Caputo is the same as that of the Riemann-Liouville derivatives.\\

 \textbf{Definition 6.} The relationship between left and right-derivatives in real space reads

\begin{equation}
D_{x^{-}}^\alpha u\left( {x} \right)=
D_{\left(-x \right)^{+}}^\alpha u\left( {x} \right)
\end{equation}
if $u(x)$ is a real and even function, and
\begin{equation}
D_{x^{-}}^\alpha u\left( {x} \right)=  -D_{\left(-x \right)^{+}}^\alpha u\left( {x} \right)
 \label{AppendixA21}
\end{equation}
if $u(x)$ is a pure imaginary and odd function.
Notice that the above transformation is valid only for $x<0$.
\\

 \textbf{Definition 7.} The modified Riemann-Liouville derivative
 reads

\begin{equation}
D_x^\alpha  u\left( {x} \right) = \frac{1} {{\Gamma \left( {1
- \alpha } \right)}}\frac{d} {{dx}}\int\limits_0^x {\left( {x - \xi
} \,\, \right)^{ - \alpha } } \left( {u\left( {\xi } \right) -
u\left( {0} \right)} \right)d\xi,
 \label{AppendixA22}
\end{equation}
where $u$ is a continuous (but not necessarily differentiable)
function. However, there exists a non-commutative property
\begin{equation}
D^{\alpha  + \beta }  \ne D^\alpha  D^\beta   \ne D^\beta  D^\alpha
\label{AppendixA23}.
\end{equation}

The Riemann-Liouville fractional derivative has some notable disadvantages
in applications such as the nonzero value 
of the fractional derivative of constants,
\begin{equation}
D_t^\alpha C= \frac{t^{-\alpha}}{\Gamma(1-\alpha)}C.
\label{AppendixA24}
\end{equation}
The Caputo fractional differentiation of a constant
results in zero. Therefore, in order to define the usual initial
value problem
\begin{equation}
u(t_0)=t_0,\,\,\,\,\,\,\,\, \left( D_t^k f \right) (t_0)=C_k ,\,\,\,\,\,\,\,\,\,\,\,\, k=1,...,n,
\label{AppendixA25}
\end{equation}
one is lead to the application of Caputo fractional derivatives instead of the Riemman-Liouville derivative.
\\
Finally, we should clearly emphasize that many usual properties of
the ordinary derivative $ D^n $ are not realized for fractional
derivative operators $ D^{\alpha} $. For example, the Leibniz rule,
chain rule, semi-group property $\left(  D_x^\alpha D_x^\alpha \ne
D_x^{2\alpha} \right)$ have strongly complicated analogs for the
operators $ D^{\alpha} $. We refer to
\cite{Podluny1999,Yang2010,Samko1993} for more informations on the
mathematical properties of fractional derivatives and integrals.

\section{}
\label{AppendixB}

Substituing Eq.~(\ref{continuumalpha1}) into Eq.~(\ref{alpha2}), one gets
\begin{equation}
\sum\limits_{\scriptstyle m =  - \infty  \hfill \atop
  \scriptstyle m \ne n \hfill}^{+\infty } {\frac{{u_n  - u_m }}{{\left| {a\left( {n - m} \right)}
\right|^{s_1} }}}  =  \frac{1}{{2\pi }}\int\limits_{ -\pi}^{
+\pi} {dk}\sum\limits_{\scriptstyle m =  - \infty  \hfill \atop
  \scriptstyle m-n \ne 0 \hfill}^{+\infty } {\frac{1}{{\left| {a\left( {n - m} \right)}
\right|^{s_1} }}}{e^{ikn} \hat u\left( {k,t} \right)} -
\frac{1}{{2\pi }}\int\limits_{ -\pi}^{ +\pi}
{dk}\sum\limits_{\scriptstyle m =  - \infty  \hfill \atop
  \scriptstyle m-n \ne 0 \hfill}^{+\infty } {\frac{1}{{\left| {a\left( {n - m} \right)}
\right|^{s_1} }}}{e^{ikm} \hat u\left( {k,t} \right)}
\label{continuumalpha1_1},
\end{equation}

\begin{equation}
\sum\limits_{\begin{array}{*{20}c}
   {m =  - \infty }  \\
   {m \ne n}  \\
 \end{array} }^{ + \infty } {\frac{{\left( {u_n  - u_m } \right)^2 }}
{{\left| {a\left( {n - m} \right)} \right|^{s_2} }}} f_{n,m}  =
\sum\limits_{\begin{array}{*{20}c}
   {m =  - \infty }  \\
 \end{array} }^{n-1} {\frac{{\left( {u_n  - u_m } \right)^2 }}
{{\left| {a\left( {n - m} \right)} \right|^{s_2} }}}  -
\sum\limits_{\begin{array}{*{20}c}
   {m = n+1}  \\
 \end{array} }^{ + \infty } {\frac{{\left( {u_n  - u_m } \right)^2 }}
{{\left| {a\left( {n - m} \right)} \right|^{s_2} }}}
 \label{continuumalpha2_1}.
\end{equation}

Denoting $n'=m-n$, one obtains for the two sums on the r.h.s. of 
Eq.~(\ref{continuumalpha2_1})

\begin{equation}
\sum\limits_{\begin{array}{*{20}c}
   {m = - \infty}  \\

 \end{array} }^{ n-1 } {\frac{{\left( {u_n  - u_m } \right)^2 }}
{{\left| {a\left( {n - m} \right)} \right|^{s_2} }}} =  \frac{{u_n
}} {{2\pi }}\int\limits_{ - \pi }^{ +\pi } {dk\left(
{\sum\limits_{n' = 1}^{ + \infty } {\frac{{1 - 2e^{ - ikn'}   }} {{\left| {an'} \right|^{s_2} }}} }
\right)} e^{ikn} \hat u\left( {k,t} \right)+  \sum\limits_{n' = 1}^{ + \infty }{\frac{u_{n-n'}}{2\pi}}\int\limits_{ -\pi }^{ + \pi}{dk\,\,{\frac{{ e^{ - ikn'}   }} {{\left| {an'} \right|^{s_2} }}}\,\,e^{ikn}\,\, \hat u\left( {k,t} \right)    }       
 \label{continuumalpha02_2},
\end{equation}
and doing the approximation that the  $u_{n \pm n'}$'s are slowly varying in space in the continuum limit, one brings the terms $u_{n \pm n'}$ outside  the sums over $n'$. Eq.~(\ref{continuumalpha02_2}) can be then rewritten as

\begin{equation}
\sum\limits_{\begin{array}{*{20}c}
   {m = - \infty}  \\

 \end{array} }^{ n-1 } {\frac{{\left( {u_n  - u_m } \right)^2 }}
{{\left| {a\left( {n - m} \right)} \right|^{s_2} }}} \approx  \frac{{u_n
}} {{2\pi }}\int\limits_{ - \pi }^{ + \pi } {dk\left(
{\sum\limits_{n' = 1}^{ + \infty } {\frac{{1 - e^{ - ikn'}   }} {{\left| {an'} \right|^{s_2} }}} }
\right)} e^{ikn} \hat u\left( {k,t} \right)
 \label{continuumalpha2_2},
\end{equation}

and

\begin{equation}
\sum\limits_{\begin{array}{*{20}c}
   {m = n+1}  \\

 \end{array} }^{ + \infty } {\frac{{\left( {u_n  - u_m } \right)^2 }}
{{\left| {a\left( {n - m} \right)} \right|^{s_2} }}}  \approx  \frac{{u_n
}} {{2\pi }}\int\limits_{ - \pi }^{ + \pi} {dk\left(
{\sum\limits_{n' = 1}^{ + \infty } {\frac{{1 - e^{ikn'} }} {{\left| {an'} \right|^{s_2} }}} }
\right)} e^{ikn} \hat u\left( {k,t} \right).
 \label{continuumalpha2_3}
\end{equation}

Hence,

\begin{equation}
\sum\limits_{\scriptstyle m =  - \infty  \hfill \atop
  \scriptstyle m \ne n \hfill}^{+\infty } {\frac{{u_n  - u_m }}{{\left| {a\left( {n - m} \right)}
\right|^{s_1} }}}  =  \frac{1}{{2\pi }}\int\limits_{ - \pi}^{
+\pi} {dk}\left(\sum\limits_{\scriptstyle n' =  - \infty  \hfill
\atop
  \scriptstyle n' \ne 0 \hfill}^{+\infty } {\frac{1}{{\left| {a{n'} }
\right|^{s_1} }}}\right){e^{ikn} \hat u\left( {k,t} \right)} -
\frac{1}{{2\pi }}\int\limits_{ - \pi}^{+\pi}
{dk}\left(\sum\limits_{\scriptstyle n' =  - \infty  \hfill \atop
  \scriptstyle n' \ne 0 \hfill}^{+\infty } {\frac{e^{ikn'}}{{\left| {a{n'} }
\right|^{s_1} }}}\right){e^{ikn} \hat u\left( {k,t} \right)}
\label{continuumalpha1_2},
\end{equation}

\begin{equation}
\sum\limits_{\begin{array}{*{20}c}
   {m =  - \infty }  \\
   {m \ne n}  \\
 \end{array} }^{ + \infty } {\frac{{\left( {u_n  - u_m } \right)^2 }}
{{\left| {a\left( {n - m} \right)} \right|^{s_2} }}} f_{n,m} =
\frac{{u_n }} {{2\pi }}\int\limits_{ -\pi }^{ + \pi}
{dk\sum\limits_{n' = 1}^{ + \infty } {\left( {\frac{{\left( {1 -
e^{ - ikn'}  } \right) - \left(
{1 - e^{ikn'}  } \right)}}
{{\left| {an'} \right|^{s_2} }}} \right)} } e^{ikn} \hat u\left(
{k,t} \right) = 
 \label{continuumalpha2_4}
\end{equation}

\begin{equation}
=
\frac{{u_n }} {{2\pi }}\int\limits_{ - \pi }^{ + \pi}
{dk{\left(\sum\limits_{n' = 1}^{ + \infty }  {{\frac{{e^{ikn'}  - e^{ - ikn'} }} {{\left| {an'} \right|^{s_2} }}} } \right)} } e^{ikn} \hat u\left( {k,t} \right)
 \label{continuumalpha2_5}.
\end{equation}

We can rewrite Eqs.~(\ref{continuumalpha1_2}) and
(\ref{continuumalpha2_5}) as

\begin{equation}
\sum\limits_{\scriptstyle m =  - \infty  \hfill \atop
  \scriptstyle m \ne n \hfill}^{+\infty } {\frac{{u_n  - u_m }}{{\left| {a\left( {n - m} \right)}
\right|^{s_1} }}}  =  \frac{1}{{2\pi }}\int\limits_{ - \pi }^{ +
\pi } {dk \left[
{\tilde J_{1}\left( 0 \right) - \tilde J_{1}\left( k \right)}
\right] e^{ikn}\,\, \hat u\left( {k,t} \right)} 
 \label{continuumalpha1_3},
\end{equation}

and

\begin{equation}
\sum\limits_{\begin{array}{*{20}c}
   {m =  - \infty }  \\
   {m \ne n}  \\
 \end{array} }^{ + \infty } {\frac{{\left( {u_n  - u_m } \right)^2 }}
{{\left| {a\left( {n - m} \right)} \right|^{s_2} }}} f_{n,m} =
\frac{{u_n }} {{2\pi }}\int\limits_{ - \pi }^{ + \pi} {dk\,\,{\tilde
J_{2}\left( k \right)} }
\,\,e^{ikn}\,\, \hat u\left( {k,t} \right)
 \label{continuumalpha2_6},
\end{equation}

where 

\begin{equation}
\tilde J_{1}\left( k \right) = \sum\limits_{\scriptstyle n =  -
\infty \hfill \atop \scriptstyle n \ne 0 \hfill}^{+\infty }
{\frac{{e^{ikn} }}{{\left| {an} \right|^{s_1}}}}
\,\,\,\,\,\,\,\,\,\text{and}\,\,\,\,\,\,\,\,\,\,\,\tilde J_{2}\left( k
\right) =  \sum\limits_{\scriptstyle n =  1 \hfill \atop
\scriptstyle \hfill}^{+\infty } {\frac{{e^{ikn} - e^{-ikn}}}{{\left|
{an} \right|^{s_2}}}} \label{continuumalpha1_4}.
\end{equation}
One has $\sum\limits_{\scriptstyle n=  1}^{  \infty }{{\frac{{e^{+ikn}
}}{{\left| {{n }} \right|^{s}}}}}= {L_i,_s}\left({e^{+ikn} }\right)
$, where ${L_i,_s}$ is the polylogarithmic function 

\begin{equation}
{L_i,_s}\left({e^{\mu}
}\right)=\Gamma\left({1-s}\right)\left({-\mu}\right)^{s-1} +
\sum\limits_{\scriptstyle n=  0}^{  \infty }{{\frac{\zeta \left(
{s-n} \right)}{{{{n !}}}}}} \left({\mu}\right)^{n}
 \label{continuumalpha1_5}.
\end{equation}

Using the Hurwitz formula \cite{Erdelyi1953}, one can rewrite 
$\tilde J_{1}\left( k \right)$ as 
\begin{equation}
\tilde J_{1}\left( k \right)= \sum\limits_{\scriptstyle n =  1
}^{+\infty } {\frac{{e^{-ikn}+e^{ikn} }}{{\left| {an}
\right|^{s_1}}}} =a^{-s_1}\left[\Gamma\left({1-s_1}\right)
\left(\left({-ik}\right)^{s_1-1}  + \left({ik}\right)^{s_1-1}\right)
+ \sum\limits_{\scriptstyle n=  0}^{  \infty }{{\frac{\zeta \left(
{s_1-2n} \right)}{{{{(2n) !}}}}}}\left( \left({-ik}\right)^{2n} +
\left({ik}\right)^{2n}\right)\right]
 \label{continuumalpha1_6}.
\end{equation}

Since 
\begin{equation}
\left({ik}\right)^{\alpha} +\left({-ik}\right)^{\alpha}=2\left|k
\right|^{\alpha}\cos\left({\frac{{\alpha \pi}}{2} }\right)
 \label{continuumalpha1_7},
\end{equation}
then
\begin{equation}
\left({ik}\right)^{2n} +\left({-ik}\right)^{2n}=2\left(-1
\right)^{n}\left|{k}\right|^{2n}
 \label{continuumalpha1_8}.
\end{equation}
Therefore
\begin{equation}
\tilde J_{1}\left( k \right)=
2a^{-s_1}\left[\Gamma\left({1-s_1}\right) \cos\left({\frac{{{s_1} -
1}}{2}\pi }\right)\left|k \right|^{s_1-1}+ \sum\limits_{\scriptstyle
n=  0}^{  \infty }{{\frac{\zeta \left( {s_1-2n} \right)}{{{{(2n)
!}}}}}}(-1)^{n}\left|{k}\right|^{2n}\right]
 \label{continuumalpha1_9}.
\end{equation}

Using the relation
\begin{equation}
\Gamma\left({1-s}\right) \cos\left({\frac{{{s} - 1}}{2}\pi }\right)=-\frac{\pi}{2\Gamma(s)\sin\left({\frac{{{s} - 1}}{2}\pi }\right)},
\label{aggiunta}
\end{equation} one gets
\begin{equation}
\tilde J_{1}\left( k \right)= 2a^{-s_1}\left[
-\frac{\pi}{2\Gamma(s_1)\sin\left({\frac{{{s_1} - 1}}{2}\pi
}\right)}\left|k \right|^{s_1-1} + \sum\limits_{\scriptstyle n=
0}^{ \infty }{{\frac{\zeta \left( {s_1-2n} \right)}{{{{(2n)
!}}}}}}(-1)^{n}\left|{k}\right|^{2n}\right]
 \label{continuumalpha1_10},
\end{equation}

and

\begin{equation}
\tilde J_{1}\left( 0\right) - \tilde J_{1}\left( k \right) =
\frac{\pi{a^{-s_1}}}{\Gamma(s_1)\sin\left({\frac{{{s_1} - 1}}{2}\pi
}\right)}\left|k \right|^{s_1-1} -2a^{-s_1}
\sum\limits_{\scriptstyle n= 0}^{  \infty }{{\frac{\zeta \left(
{s_1-2n} \right)}{{{{(2n) !}}}}}}(-1)^{n}\left|{k}\right|^{2n} +
2a^{-s_1} \zeta \left( {s_1}\right)
 \label{continuumalpha1_11}.
\end{equation}

We observe that:

\begin{equation}
\tilde J_{1}\left( 0\right) - \tilde J_{1}\left( k \right)  \sim \left\{ {\begin{array}{*{20}c}
 \left|k \right|^{s_1-1},  \,\, 1< s_1 <3 , \,\, s_1 \ne  2 \\
 \\
  - \zeta \left({s_1-2} \right)k^2 , \,\, s_1>3 \\
\end{array}} \right.
 \label{continuumalpha1_110}.
\end{equation}

The connection between the Riesz fractional derivative and its Fourier transform \cite{Samko1993} is given by:

\begin{equation}
\left |k\right |^\alpha  \longleftrightarrow -\frac{{\partial ^\alpha }} {{\partial \left |x\right |^\alpha }}, \,\,\,\,\,\, k^2  \longleftrightarrow -\frac{{\partial ^2 }} {{\partial \left |x\right |^2 }} 
 \label{continuumalpha1_1100}.
\end{equation}

Then, when $s>3$ we get back the conventional short-range $\alpha$-FPU model.
It is easy to observe that $ \tilde J_1 \left( 0 \right) - \tilde
J_1 \left( k \right) \sim \omega^2 \left(k\right) $. In the same
way, i.e. combining the polylogarithmic function and the Hurwitz
formula, we obtain

\begin{equation}
 \label{continuumalpha2_7} 
J_2\left( k \right) =a^{ - s_2} \Gamma \left( {1 -
s_2} \right)\left[ {\left( { - ik} \right)^{s_2 - 1}  - \left( {ik}
\right)^{s_2 - 1} } \right] + a^{ - s_2} \sum\limits_{n = 0}^{ +
\infty } {\frac{{\zeta \left( {s_2
- n} \right)}} {{n!}}} \left[ {\left( { - ik} \right)^n  - \left(
{ik} \right)^n } \right]. 
\end{equation}

 In Eqs.~(\ref{continuumalpha1_11}) and (\ref{continuumalpha2_7}),
$\zeta$ is the Riemann zeta function. In the continuum limit $
\left( {k \to 0} \right)$ the long-wavelength modes are singled out
and the leading terms of Eqs.~(\ref{continuumalpha1_3}) and
(\ref{continuumalpha2_6}) are derived from the first terms of the
r.h.s. of Eqs.~(\ref{continuumalpha1_11}) and
(\ref{continuumalpha2_7}). Therefore,

\begin{equation}
\sum\limits_{\scriptstyle m =  - \infty  \hfill \atop
  \scriptstyle m \ne n \hfill}^{+\infty } {\frac{{u_n  - u_m }}{{\left| {a\left( {n - m} \right)} \right|^{s_1} }}} \approx 
\frac{1}{{2\pi }}\int\limits_{ - \pi }^{ +\pi } {dk \,  \left( \frac{\pi{a^{1-s_1}}}{\Gamma(s_1)\sin\left({\frac{{{s_1} - 1}}{2}\pi }\right)}\left|k \right|^{s_1-1}\right) e^{ikn}\,\, \hat u\left( {k,t}
  \right)}.
\label{continuumalpha1_12}
\end{equation}

\begin{equation}
\sum\limits_{\begin{array}{*{20}c}
   {m =  - \infty }  \\
   {m \ne n}  \\
 \end{array} }^{ + \infty } {\frac{{\left( {u_n  - u_m } \right)^2 }}
{{\left| {a\left( {n - m} \right)} \right|^{s_2} }}} f_{n,m} \approx
\frac{{u_n }} {{2\pi }}\int\limits_{ - \pi }^{ + \pi }
{dk\left\{ {a^{1 - s_2} \Gamma \left( {1 - s_2} \right)\left( {\left( { - ik} \right)^{s_2 - 1}  -
\left( {ik} \right)^{s_2 - 1} } \right)} \right\}} e^{ikn} \hat
u\left( {k,t} \right)
 \label{continuumalpha2_8}.
\end{equation}

Taking into account that $n=x_n/a$, in the continuum limit $a \to 0$ 
one can replace $n$ by $x/a$, where $x$ now labels a point in the real line. 
In this limit, using Eq.~(\ref{aggiunta}) and 
writing explicitly the lattice spacing $a$,  
we obtain

\begin{equation}
\sum\limits_{\scriptstyle m =  - \infty  \hfill \atop
  \scriptstyle m \ne n \hfill}^{+\infty } {\frac{{u_n  - u_m }}{{\left| {a\left( {n - m} \right)} \right|^{s_1} }}} \approx 
\frac{1}{{2\pi }}\int\limits_{ - \infty }^{ +\infty } {\frac{{dk}} {a} \, e^{ik\frac{x}{a}} \left( \frac{\pi}{\Gamma(s_1)\sin\left({\frac{{{s_1} - 1}}{2}\pi }\right)}\right)\left|\frac{k}{a} \right|^{s_1-1} a\,\tilde u\left( {k,t} \right)}.
\label{continuumalpha1_13}
\end{equation}

\begin{equation}
\sum\limits_{\begin{array}{*{20}c}
   {m =  - \infty }  \\
   {m \ne n}  \\
 \end{array} }^{ + \infty } {\frac{{\left( {u_n  - u_m } \right)^2 }}
{{\left| {a\left( {n - m} \right)} \right|^{s_2} }}} f_{n,m} \approx
\frac{a\,{u\left( {x,t} \right) }} {{2\pi }}\int\limits_{ - \infty }^{ + \infty }
{\frac{{dk}} {a}\left\{ {\frac{{\pi }} {{\Gamma \left( s_2 \right)\sin \left( {s_2\pi }
\right)}}\left( {\left( { - i\frac{k}{a}} \right)^{s_2 - 1}  -
\left( {i\frac{k}{a}} \right)^{s_2 - 1} } \right)} \right\}} e^{ikn}\,
a\,\tilde u\left( {k,t} \right)
 \label{continuumalpha2_9}.
\end{equation}

Setting ${k}/{a}=p$, it follows
\begin{equation}
\sum\limits_{\scriptstyle m =  - \infty  \hfill \atop
  \scriptstyle m \ne n \hfill}^{+\infty } {\frac{{u_n  - u_m }}{{\left| {a\left( {n - m} \right)} \right|^{s_1} }}}  \approx 
  \frac{\pi }{{\Gamma \left( s \right)\sin \frac{{{s_1} - 1}}{2}\pi }}\frac{1}{{2\pi }}\int\limits_{ - \infty }^{ +\infty } {dp\,\,e^{ipx} \left| p \right|^{s_1 - 1} \hat u\left( {p,t} \right)}
 \label{continuumalpha1_141},
\end{equation}

\begin{equation}
\sum\limits_{\begin{array}{*{20}c}
   {m =  - \infty }  \\
   {m \ne n}  \\
 \end{array} }^{ + \infty } {\frac{{\left( {u_n  - u_m } \right)^2 }}
{{\left| {a\left( {n - m} \right)} \right|^{s_2} }}} f_{n,m} \approx
\frac{{a\,u\left( {x,t} \right)}} {{2\pi }}\int\limits_{ - \infty }^{ +
\infty } {dp\,\,e^{ipx} \frac{{\pi
}} {{\Gamma \left( s_2 \right)\sin \left( {s_2\pi } \right)}}\left(
{\left( { - ip} \right)^{s_2 - 1}  - \left( {ip} \right)^{s_2 - 1} }
\right)} \hat u\left( {p,t} \right),
 \label{continuumalpha2_101}
\end{equation}
which are the correct expressions in the continuum limit. 
In the previous 
Eqs.(\ref{continuumalpha1_13})-(\ref{continuumalpha2_101}) 
we replaced the discrete function $u_n(t)$ 
according to the relation:
\begin{equation}
u_{n}\left(t\right) \equiv a\,u\left(x_n,t\right)
 \label{continuumalpha2_102}.
\end{equation}
We assumed \cite{Tarasov2011} 
\begin{equation}
 \tilde u\left( {k,t} \right) =\mathcal{L} \hat u\left( {k,t} \right)
 \label{continuumalpha2_102_bis},
\end{equation}
where $\mathcal{L}$ denotes the limit $a\to 0$. Note that $\tilde u\left( {k,t} \right)$ is the Fourier transform of the field $u\left( {x,t} \right)$, 
and $\hat u\left( {p,t} \right) \equiv a \tilde u\left( {k,t} \right)$.

\section{}
\label{AppendixC}

One has

\begin{widetext}
\begin{equation}
\sum\limits_{\begin{array}{*{20}c}
   {m =  - \infty }  \\
   {m \ne n}  \\
 \end{array} }^{m =  + \infty } {\frac{{\left( {u_n  - u_m } \right)^2 }}
{{\left| {a\left( {n - m} \right)} \right|^s }}}  =
\sum\limits_{\begin{array}{*{20}c}
   {m =  - \infty }  \\
   {m \ne n}  \\
 \end{array} }^{m =  + \infty } {\frac{{u_n^2  - 2u_n u_m  \,\, {+}\,\, u_m^2 }}
{{\left| {a\left( {n - m} \right)} \right|^s }}}=
 \label{alternatingalpha4}
\end{equation}
\end{widetext}

\begin{equation}
\begin{array}{l}
\sum\limits_{\begin{array}{*{20}c}
   {m = - \infty}  \\
   {m - n \ne 0}  \\
 \end{array} }^{ + \infty } {\frac{{\left( {u_n  - u_m } \right)^2 }}
{{\left| {a\left( {n - m} \right)} \right|^{s_2} }}} =  \frac{{u_n
}} {{2\pi }}\int\limits_{ - \pi }^{ +\pi } {dk\left(
{\sum\limits_{n' = 1}^{ + \infty } {\frac{{2 - 2\left(e^{ - ikn'} + e^{ - ikn'} \right)  }} {{\left| {an'} \right|^{s_2} }}} }
\right)} \,\,e^{ikn} \,\, \hat u\left( {k,t} \right) + 
\frac{1} {{2\pi }}  \sum\limits_{n' = 1}^{ + \infty }{u_{n-n'}}\int\limits_{ - \pi}^{ +\pi}{dk\,\,{\frac{{ e^{ - ikn'}   }} {{\left| {an'} \right|^{s_2} }}}\,\,e^{ikn} \,\,\hat u\left( {k,t} \right) } \,\, + \\  
+\frac{1} {{2\pi }}  \sum\limits_{n' = 1}^{ + \infty }{u_{n+n'}}\int\limits_{ - \pi }^{ + \pi }{dk\,\,{\frac{{ e^{ + ikn'}   }} {{\left| {an'} \right|^{s_2} }}}\,\,e^{ikn} \,\,\hat u\left( {k,t} \right) }
 \end{array}
 \label{alternatingalpha04}.
\end{equation}

Doing the approximation that the $u_{n \pm n'}$'s are slowly varying in space in the continuum limit, 
Eq.~(\ref{alternatingalpha04}) can be rewritten as

\begin{widetext}
\begin{equation}
\sum\limits_{\begin{array}{*{20}c}
   {m =  - \infty }  \\
   {m \ne n}  \\
 \end{array} }^{m =  + \infty } {\frac{{\left( {u_n  - u_m } \right)^2 }}
{{\left| {a\left( {n - m} \right)} \right|^s }}}  
= \frac{{u_n }}
{{2\pi }}\int\limits_{ - \pi }^{ + \pi } {dk}
\sum\limits_{\begin{array}{*{20}c}
   {n' =  1}  \\
 \end{array} }^{n' =  + \infty } {\left( {\frac{2}
{{\left| {an'} \right|^s }} - \frac{{e^{-ikn'} +e^{ikn'} }} {{\left| {an'}
\right|^s }} } \right)} e^{ikn} \hat u\left( {k,t} \right)=
 \label{alternatingalpha5}
\end{equation}
\end{widetext}

\begin{widetext}
\begin{equation}
%
= \frac{{u_n }}
{{2\pi }}\int\limits_{ - \pi }^{ +\pi } {dk}\left[ {\tilde J\left( 0 \right) - \tilde J\left( k \right) } \right] e^{ikn}\,\,
\hat u\left( {k,t} \right)
 \label{alternatingalpha6},
\end{equation}
\end{widetext}

where $\tilde J\left( k \right) = \sum\limits_{\scriptstyle n =  -
\infty \hfill \atop \scriptstyle n \ne 0 \hfill}^{+\infty }
{\frac{{e^{ikn} }}{{\left| {an} \right|^{s}}}}$. 
Using again the Hurwitz formula, 
$\tilde J\left( 0 \right) - \tilde J\left( k \right) $ is rewritten as
\begin{equation}
 \label{alteratingalpha8} J\left( 0 \right) - \tilde J\left( k \right) = \frac{{\pi a^{ - s} }} {{\Gamma \left( s \right)\sin \left(
{\frac{{s - 1}} {2}\pi } \right)}}\left| k \right|^{s - 1} - a^{ - s}
\sum\limits_{n = 0}^{ + \infty } {\frac{{\zeta \left( {s - n}
\right)}} {{n!}}}  \left[
{\left(-ik \right)^{n}  +\left(+ik \right)^{n}   } \right]
 + 2a^{ - s} \zeta \left( s \right). 
\end{equation}

In the continuum limit $ \left( {k \to 0} \right)$, the leading term
of Eq.~(\ref{alternatingalpha6}) is derived from the first term of
the r.h.s. of Eq.~(\ref{alteratingalpha8}):

\begin{widetext}
\begin{equation}
\sum\limits_{\begin{array}{*{20}c}
   {m =  - \infty }  \\
   {m \ne n}  \\
 \end{array} }^{m =  + \infty } {\frac{{\left( {u_n  - u_m } \right)^2 }}
{{\left| {a\left( {n - m} \right)} \right|^s }}}  \approx \frac{{u\left( {x,t} \right)
}} {{2\pi }}\int\limits_{ - \infty }^{ + \infty } {dk}\,\,
e^{i k x} \left\{ {\frac{{\pi a^{1 - s} }} {{\Gamma \left( s
\right)\sin \left( {\frac{{s - 1}} {2}\pi } \right)}}\left|
k \right|^{s - 1}}
\right\}\tilde u\left( {k,t} \right) .
 \label{alteratingalpha9}
\end{equation}
\end{widetext}



\section{}
\label{AppendixD}

We start by observing

\begin{widetext}
\begin{equation}
\sum\limits_{\begin{array}{*{20}c}
   {m =  - \infty }  \\
   {m \ne n}  \\
 \end{array} }^{m =  + \infty } {\frac{{\left( {u_n  - u_m } \right)^3 }}
{{\left| {a\left( {n - m} \right)} \right|^s }}}  =
\sum\limits_{\begin{array}{*{20}c}
   {m =  - \infty }  \\
   {m \ne n}  \\
 \end{array} }^{m =  + \infty } {\frac{{u_n^3  - 3u_n^2 u_m  + 3u_n u_m^2  - u_m^3 }}
{{\left| {a\left( {n - m} \right)} \right|^s }}}=
 \label{beta3}
\end{equation}
\end{widetext}

\begin{widetext}
\begin{equation}
\begin{array}{l}
= \frac{{u_n^2
}} {{2\pi }}\int\limits_{ -\pi }^{ + \pi } {dk}
\sum\limits_{\begin{array}{*{20}c}
   {n' =  1 }  \\
 \end{array} }^{+ \infty } {{
 \frac{{2-3\left(e^{-ikn'} +e^{ikn'}\right)}} {{\left| {an'}
\right|^{s_2} }} } }\,\, e^{ikn}\,\,
\hat u\left( {k,t} \right) +  \sum\limits_{n' = 1}^{ + \infty }{ \frac{u_{n}u_{n-n'}} {{2\pi }} }\int\limits_{ - \pi }^{ + \pi }{dk\,\,{\frac{{ 3e^{ - ikn'}   }} {{\left| {an'} \right|^{s_2} }}}\,\,e^{ikn} \,\,\hat u\left( {k,t} \right) }\,\, + \\
+\sum\limits_{n' = 1}^{ + \infty }{ \frac{u_{n}u_{n
 +n'}} {{2\pi }} }\int\limits_{ -\pi }^{ + \pi }{dk\,\,{\frac{{ 3e^{ + ikn'}   }} {{\left| {an'} \right|^{s_2} }}}\,\,e^{ikn} \,\,\hat u\left( {k,t} \right) } -  \sum\limits_{n' = 1}^{ + \infty }{ \frac{u_{n
 -n'}^2} {{2\pi }} }\int\limits_{ - \pi }^{ + \pi }{dk\,\,{\frac{{ e^{ - ikn'}   }} {{\left| {an'} \right|^{s_2} }}}\,\,e^{ikn} \,\,\hat u\left( {k,t} \right) } -  \sum\limits_{n' = 1}^{ + \infty }{ \frac{u_{n
 +n'}^2} {{2\pi }} }\int\limits_{ -\pi }^{ +\pi }{dk\,\,{\frac{{ e^{ + ikn'}   }} {{\left| {an'} \right|^{s_2} }}}\,\,e^{ikn} \,\,\hat u\left( {k,t} \right) }

 \end{array}
 \label{beta8}.
\end{equation}
\end{widetext}

Again doing the approximation that the $u_{n \pm n'}$'s are slowly varying 
Eq.~(\ref{beta8}) can be rewritten as

\begin{widetext}
\begin{equation}
\sum\limits_{\begin{array}{*{20}c}
   {m =  - \infty }  \\
   {m \ne n}  \\
 \end{array} }^{m =  + \infty } {\frac{{\left( {u_n  - u_m } \right)^3 }}
{{\left| {a\left( {n - m} \right)} \right|^s }}}  
= \frac{{u_{n}^2 }}
{{2\pi }}\int\limits_{ - \pi}^{ + \pi} {dk}
\sum\limits_{\begin{array}{*{20}c}
   {n' =  1}  \\
 \end{array} }^{n' =  + \infty } {\left( { \frac{{2-\left(e^{-ikn'} +e^{ikn'} \right)}} {{\left| {an'}
\right|^s }} } \right)} e^{ikn} \hat u\left( {k,t} \right)=
 \label{beta00}
\end{equation}
\end{widetext}

\begin{widetext}
\begin{equation}
%
= \frac{{u_{n}^2 }}
{{2\pi }}\int\limits_{ - \pi }^{ + \pi } {dk}\left[ {\tilde J\left( 0 \right) - \tilde J\left( k \right) } \right] e^{ikn}\,\,
\hat u\left( {k,t}
 \right)
 \label{beta4}.
\end{equation}
\end{widetext}

Once again, combining the polylogarithmic function and the Hurwitz
formula, we obtain
\begin{equation}
 \label{alteratingalpha8_bis} J\left( 0 \right) - \tilde J\left( k \right) = \frac{{\pi a^{ - s_2} }} {{\Gamma \left( s_2 \right)\sin \left(
{\frac{{{s_2} - 1}} {2}\pi } \right)}}\left| k \right|^{{s_2} - 1} - a^{ - s_2}
\sum\limits_{n = 0}^{n =  + \infty } {\frac{{\zeta \left( {s_2 - n}
\right)}} {{n!}}}  \left[
{\left(-ik \right)^{n}  +\left(+ik \right)^{n}   } \right]
 + 2a^{ - s_2} \zeta \left( s_2 \right). 
\end{equation}

Therefore, in the continuum limit

\begin{widetext}
\begin{equation}
\sum\limits_{\begin{array}{*{20}c}
   {m =  - \infty }  \\
   {m \ne n}  \\
 \end{array} }^{m =  + \infty } {\frac{{\left( {u_n  - u_m } \right)^3 }}
{{\left| {a\left( {n - m} \right)} \right|^{s_2} }}}  \approx
\frac{{u^2 \left( {x,t} \right) }} {{2\pi }}\int\limits_{ - \infty }^{ + \infty }
{dk}\,\, e^{i k x} \left\{ {\frac{{\pi a^{1 - s_2} }} {{\Gamma
\left( s_2 \right)\sin \left( {\frac{{s_2 - 1}} {2}\pi } \right)}}\left| k \right|^{s_2 - 1}  } \right\}\tilde
u\left( {k,t} \right).
 \label{beta7}
\end{equation}
\end{widetext}

%
%

\section{The ($\alpha$+$\beta$)--FPU model with
power-law long-range interactions} \label{AppendixE}

For the sake of completeness, we report in this Appendix results for
the $\alpha$+$\beta$--FPU model in which cubic and quartic terms are both
present. The Hamiltonian of the model reads

\begin{widetext}
\begin{equation}
H  = \frac{1}{2}\sum\limits_{n =  -
\infty }^{+\infty } {\dot u_n^2 } + \frac{\chi
}{4}\sum\limits_{\begin{array}{*{20}c}
   {n,m =  - \infty }  \\
   {m{\ne}n}  \\
\end{array}}^{+\infty } {\frac{{\left[ {u_n  - u_m } \right]^2  }}{{\left| {a\left( {n - m} \right)} \right|^s }}}
+ \frac{\gamma }{3}\sum\limits_{\begin{array}{*{20}c} {n,m =  - \infty }  \\
   {m{\rm{ < }}n}  \\
\end{array}}^{+\infty } {\frac{{\left[ {u_n  - u_m } \right]^3 }}{{\left| {a\left( {n - m} \right)} \right|^s }}} +
\frac{\lambda }{8}\sum\limits_{\begin{array}{*{20}c}
   {n,m =  - \infty }  \\
   {m{\ne}n}  \\
\end{array}}^{+\infty} {\frac{{\left[ {u_n  - u_m } \right]^4 }}{{\left| {a\left( {n - m} \right)} \right|^s }}}
 \label{alpha+beta1}
\end{equation}
\end{widetext}

One gets

\begin{widetext}
\begin{equation}
\ddot u_n  + \chi \sum\limits_{\begin{array}{*{20}c}
   {m =  - \infty }  \\
   {m \ne n}  \\
\end{array}}^{+\infty } {\frac{{u_n  - u_m }}{{\left| {a\left( {n - m} \right)} \right|^s }}}
              + \gamma \sum\limits_{\begin{array}{*{20}c}{m =  - \infty }  \\
   {m \ne n}  \\
\end{array}}^{+\infty } {\frac{{\left[ {u_n  - u_m } \right]^2 }}{{\left| {a\left( {n - m} \right)} \right|^s }}}{f_n,_m}
+ \lambda \sum\limits_{\begin{array}{*{20}c}{m =  - \infty }  \\
   {m \ne n}  \\
\end{array}}^{+\infty } {\frac{{\left[ {u_n  - u_m } \right]^3 }}{{\left| {a\left( {n - m} \right)} \right|^s }}}=0,
 \label{alpha+beta2}
\end{equation}
\end{widetext}
where again
$$ {f_n, _m}  = \left\{ {\begin{array}{*{20}c}
   +1 & {,m < n}  \\
   -1 & {,m > n.}  \\
\end{array}} \right.$$ \\

In the continuum limit for a lattice field model, introducing
Eqs.~(\ref{continuumalpha1_15}), (\ref{continuumalpha2_11}) and (\ref{beta11}) into Eq.~(\ref{alpha+beta2}), we get
\begin{widetext}
\begin{equation}
\frac{{\partial ^2 }} {{\partial t^2 }}u\left( {x,t} \right) - g_{s
- 1} \frac{{\partial ^{s - 1} }} {{\partial \left| x \right|^{s - 1}
}}u\left( {x,t} \right) - h_{s - 1} u\left( {x,t} \right)\left[
{D_{x^ -  }^{s - 1}  - D_{x^ +  }^{s - 1} } \right]u\left( {x,t}
\right) - r_{s - 1} u^2 \left( {x,t} \right)\frac{{\partial ^{s - 1}
}} {{\partial \left| x \right|^{s - 1} }}u\left( {x,t} \right) = 0,
 \label{alpha+beta3}
\end{equation}
\end{widetext}
where the constants $g_{{s} - 1}$ and $k_{{s} - 1}$
are given by Eq.~(\ref{continuumlimitalpha17}) and Eq.~(\ref{alternatingalpha12}), respectively (when ${s_2}={s_1}={s}$). $r_{{s} - 1}  =j_{{s} - 1}\left(\gamma=\lambda \right)$ and $j_{{s} - 1}$ is given by Eq.~(\ref{beta13}).

\end{document}